# Heterostructures produced from nanosheet-based inks

F. Withers[1], H. Yang[2], L. Britnell[1], A.P Rooney[3], E. Lewis[3], A. Felten[4], C. R. Woods[1], V. Sanchez Romaguera[2], T. Georgiou[1], A. Eckmann[2], Y.J. Kim[1,5], S. G. Yeates[2], S. J. Haigh[3], A. K. Geim[6], K. S. Novoselov[1], C. Casiraghi[2]*

[1]School of Physics and Astronomy, University of Manchester, Oxford Road, Manchester, M13 9PL, UK

[2]School of Chemistry, University of Manchester, Oxford Road, Manchester, M13 9PL, UK

[3]School of Materials, University of Manchester, Manchester M13 9PL, UK

[4]Research Centre in Physics of Matter and Radiation (PMR), Université de Namur, B-5000 Namur, Belgique

[5]Department of Chemistry, College of Natural Sciences, Seoul National University, Seoul, 151-747, Korea

[4]Manchester Centre for Mesoscience and Nanotechnology, University of Manchester, Oxford Road, Manchester, M13 9PL, UK

[*]Corresponding author: cinzia.casiraghi@manchester.ac.uk

**ABSTRACT**

**The new paradigm of heterostructures based on two-dimensional (2D) atomic crystals has already led to the observation of exciting physical phenomena and creation of novel devices. The possibility of combining layers of different 2D materials in one stack allows unprecedented control over the electronic and optical properties of the resulting material. Still, the current method of mechanical transfer of individual 2D crystals, though allowing exceptional control over the quality of such structures and interfaces, is not scalable. Here we show that such heterostructures can be assembled from chemically exfoliated 2D crystals, allowing for low-cost and scalable methods to be used in the device fabrication.**

Keywords: graphene, 2D crystals, inks, devices, heterostructures, flexible electronics

A large variety of 2D atomic crystals isolated in the recent years offer a rich platform for the creation of heterostructures(*1-3*) which combine several of these materials in one stack. Since, collectively, this class of materials covers a very broad range of properties, the obtained heterostructures can be tuned to focus on particular phenomena, or be used for specific applications (4-15) (or even to perform multiple functions). Still, up to now all vertical heterostructures have been produced by micromechanical cleavage(16) of three-dimensional layered crystals with subsequent dry transfer(*4, 6*) of each crystal layer. While this technique allows one to achieve extremely high quality stacks(17), it certainly cannot be applied to the production of such heterostructures on a large scale. So, alternative methods, compatible with mass-production and that does not require the use of clean rooms and expensive techniques such as lithography should be utilised to bring the attractive qualities of such systems to real-life applications.

One approach is based on the use of liquid phase exfoliation (LPE) to produce dispersions of various 2D crystals (18-20). One can then use such inks to deposit platelet layers of different materials sequentially by standard low-cost fabrication techniques (drop- and spray-coatings, roll-to-roll transfer, ink-jet printing (*21*), *etc.*). LPE has already been used as a mass-scale approach for production of 2D crystals, and offers several advantages for cost reduction and scalability. One of the most important advantages of LPE is that the same method can be used to create inks made of nanosheets of different 2D crystals , covering a large variety of properties. Furthermore, this technique is compatible with low-cost and flexible substrates, so it is expected to have a big impact on the new generation of flexible electronics and photovoltaics (*11*). Also, despite the small size of the flakes and the use of the surfactants, the flakes of the 2D materials still demonstrate similar properties to those of the large scale 2D crystals obtained by micromechanical cleavage(*19-20*), thus creating stable, long-lasting inks with well-defined properties. Such LPE-produced inks and heterostructures have already been used to make simple devices such as sensors for hydrogen evolution reaction (*22*) and planar photo-voltaic devices (*23*). However, here we will show that more complex, multi-functional and flexible devices, based on completely different physical principles, can be created.

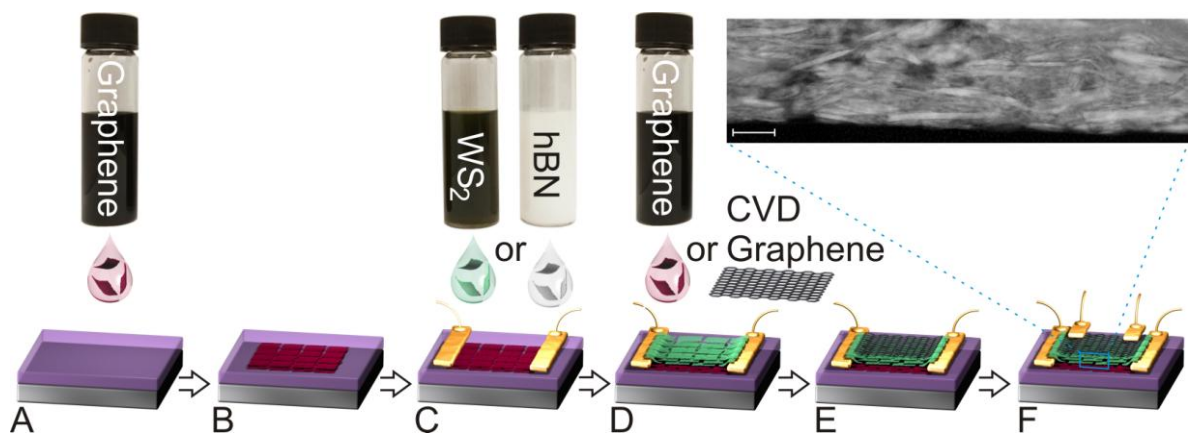

*Figure 1. Schematic of a general heterostructure device fabrication process made by using 2D-crystal inks. The dispersions in A, C and D are concentrated aqueous dispersions of (left to right) graphene, $WS_2$, h-BN and graphene. (F) The final device and the cross-sectional STEM HAADF image of the $WS_2$ thin film. The scale bar is 35 nm*

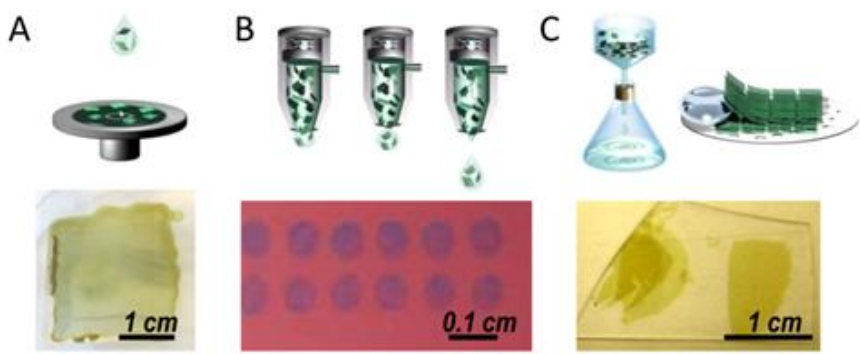

*Figure 2 Schematic representations of deposition methods for LPE 2D atomic crystals and optical micrographs of deposited films: (A) drop-casting on glass; (B) ink-jet printing on Si/SiO$_2$(300nm); (C) vacuum filtration and fishing on glass.*

The purpose of this work is to show the proof of concept of a promising low cost technology(*24*), suitable for the mass-production of devices based on heterostructures, which can be applied to make devices of arbitrary complexity. Here we take this concept to the ultimate level, combining a number of different materials in a controlled vertical stack. In particular we developed a technological platform to controllably create vertical heterostructures, which can be used to make a large variety of devices. We present several examples of such heterostructures created by depositing LPE 2D crystals via drop-casting, ink-jet printing and vacuum filtration. In particular, we make use of graphene (Gr), transition metal dichalcogenides (TMDC, such as WS$_2$ and MoS$_2$), and hexagonal-boron nitride (h-BN) inks (details on the production and characterization of the dispersions can be found in (*25*)). These crystals have been selected because of their complementary electronic and dielectric properties, ranging from the high transparency and conductivity of graphene(*26*), to the high optical absorption of TMDC(*27*) and the high transparency and resistivity of h-BN(*14,28,29*). We fabricated and tested different types of devices with the following general structure: BGr/Barrier/TGr, where TGr and BGr refer to top and bottom graphene electrodes, respectively (Fig. 1). Here we demonstrate that such devices can act as (i) tunnelling transistors(*11, 12, 3*0), where tunnelling between TGr and BGr through a barrier (typically made of h-BN or TMDC) is controlled by a back gate; (ii) photovoltaic devices where light absorbed in the barrier (TMDC) is converted into photocurrent through TGr and BGr; (iii) in-plane transistors, where TGr is used as a gate and the Barrier as a gate dielectric to control the in-plane current in BGr.

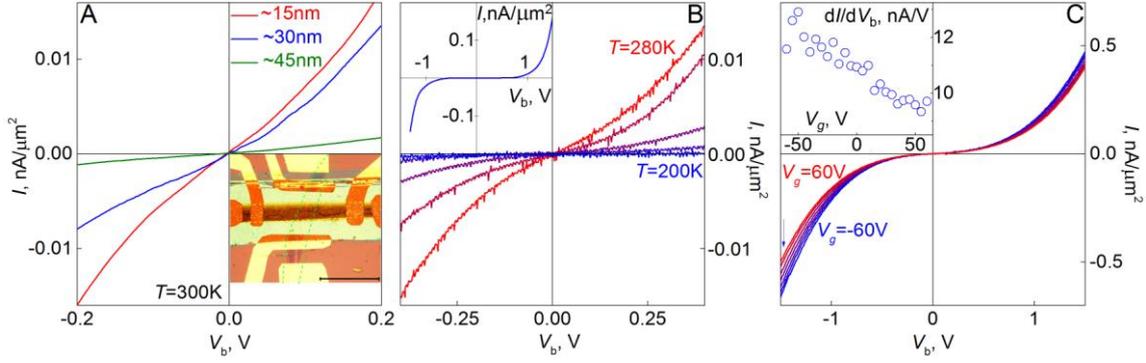

*Figure 3 (A) I-$V_b$ curves for Si/SiO$_2$/BGr/WS$_2$/TGr heterostructures with different thickness of WS$_2$. $V_g$=0V. Inset: optical micrograph of one of our devices. Boundaries of BGr (yellow, produced by drop coating) and TGr (green, mechanically exfoliated few layer graphene) are marked by dashed lines. The whitish (when on Si/SiO$_2$) or reddish (when on gold contacts) area is LPE WS$_2$. Scale bar 100μm. (B) Temperature dependence (from T=200K (blue) to T=280K (red) in 20K steps) of the I-$V_b$ characteristics of a BGr/WS$_2$/TGr device (WS$_2$ thickness ~30nm, here we used mechanically exfoliated graphene as BGr and TGr). Inset: I-$V_b$ for the same device at T=4.2K. $V_g$=0V. (C) I-$V_b$ characteristics for the same device at different $V_g$ (from $V_g$=-60V (blue) to $V_g$=60V (red) in 20V steps). T=300K. Inset: Differential conductivity of the same device at $V_b$=-1.5V (marked by the blue arrow in the main panel) as a function of $V_g$.*

RESULTS AND DISCUSSION

Fig. 1 shows a schematic of a general process used to fabricate such devices: graphene ink is deposited on a Si/SiO$_2$ substrate (Fig. 1A) to fabricate the bottom electrode (Fig. 1B). Then TMDC or h-BN inks are used to fabricate a thin film on top of the bottom electrode (Fig. 1C). Cross sectional scanning transmission electron microscopy (STEM) high angle annular dark field (HAADF) imaging has been used to access the quality and the stacking of the laminates in our inks (Fig. 1F). The TGr is usually composed of CVD (Chemical Vapour Deposition), LPE or mechanically exfoliated graphene (Fig. 1D,E) to ensure sufficient optical transparency. In the case of using mechanically exfoliated graphene, the deposition was done by using a dry transfer method(*4, 6*).

Three low cost and scalable methods have been used for the deposition of LPE 2D crystals: drop-casting, ink-jet printing, and vacuum filtration (Fig. 2). We utilised different types of dispersions (depending of the deposition method used): from N-methylpyrrolidone-based (NMP) dispersions (*19-20*) to aqueous dispersions obtained by mixed solvents (31) or by using pyrene derivatives(32). Compared to generally used NMP and DMF dispersions of graphene and other 2D crystals, such aqueous/mixed dispersions offer much faster drying rate and contains a lower amount of stabilizer, which is essential when using our deposition methods (for instance in low temperature post-printing processing). Aqueous dispersions are also much more environmentally friendly.

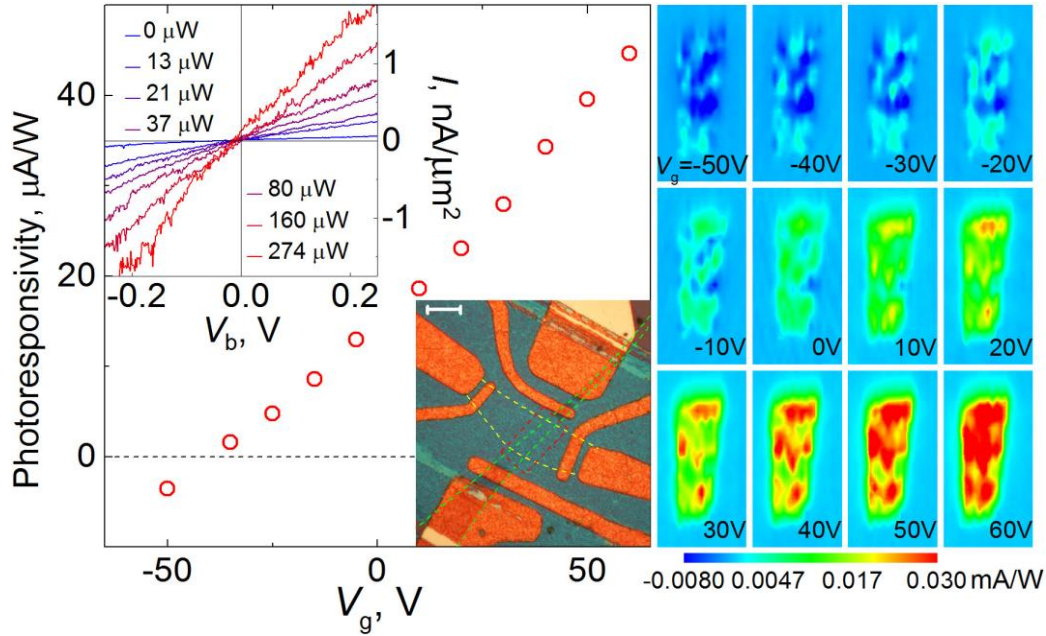

*Figure 4. Left panel: Photoresponsivity (zero-bias photocurrent) of a Si/SiO$_2$/BGr/WS$_2$/TGr device as a function of $V_g$. Each point is obtained by averaging the zero bias photocurrent maps (right panel). The photocurrent changes sign at $V_g \approx 40V$, indicating reversing in the direction of the built in electric field. Top inset: I-$V_b$ characteristics of the device at different laser powers. Bottom inset: optical micrograph of one of our devices. Boundaries of BGr (yellow) and TGr (green) are marked by dashed lines. Both BGr and TGr are produced by transferring mechanically exfoliated graphene. Greenish (when on Si/SiO$_2$) or reddish (when on gold contacts) area is LPE WS$_2$. Scale bar 10µm. Right panel: spatial maps of the zero bias photocurrent for the same device, taken from the area marked by red rectangular in the bottom insert to the left panel at different values of $V_g$. The width of each map is 10µm. Incident power 56µW, laser energy 1.96eV (See (25) for different excitation energies).*

In the specific case of vertical heterostructure the quality of the film is determined by the amount of defects and pin-holes. Therefore, it is essential to characterize and compare the coatings produced with different methods (*25*). For the purposes of our, relatively small (typically sub-mm), device, we observed that the three methods give very similar results. Raman spectroscopy does not show any strong changes between TMDC films obtained with the three methods. Furthermore, all the methods allow one to produce pinholes-free films: (i) ink-jet printing allows controlling the shape of the films (stripes, dots, *etc*) and to remove pinholes by printing several times over the same feature (*21*); (ii) drop casting and spray coating allow covering large area and pinholes free films can be obtained by using highly concentrated dispersions; (iii) filtering and fishing allows covering smaller area, as compared to drop casting, but the pinholes density can be controlled by transferring on the same area several times (typically, the density of the pinholes is strongly reduced after 2 transfers, (*25*)). The presence of pinholes can immediately be detected by very low tunnelling resistance of our tunnel junctions. At the same time, finite size and stiffness of TMDC and graphene flakes makes the device insensitive to small pin-holes due to "bridging" effect. Further details of the methods can be found in (*25*).

We start with tunnelling junctions and tunnelling transistors which have a structure of Si/SiO$_2$/BGr/WS$_2$/TGr (MoS$_2$ – based devices demonstrate similar characteristics (*25*)). Tunnelling junctions may have both BGr and TGr produced by either of the methods mentioned above, whereas tunnelling transistors require exactly monolayer graphene to be used as BGr in order not to screen the gate voltage (and as such are prepared from CVD or mechanically exfoliated graphene). The current-bias voltage (*I*-*V*$_b$) characteristics for our devices are strongly non-linear (Fig. 3). As expected, the zero-bias conductivity goes down as the thickness of WS$_2$ layer increases (Fig. 3A). The uncertainty in the thickness of the layer (RMS roughness ~3nm) prevents us from any quantitative analysis of the scaling behaviour. Zero-bias conductivity also decreases dramatically with decreasing temperature (Fig. 3B). Such a strong temperature dependence suggests an excitation mechanism for charge carrier generation, either from the graphene electrodes (in this case the tunnelling barrier is the Schottky barrier between graphene and WS$_2$) or from the impurity band in WS$_2$ (a strong impurity band is expected due to the large fraction of edges in our nanocrystals of WS$_2$). We would like to note that a variation in the tunnelling barrier thickness leads to effectively a range of tunnelling barriers connected in parallel and can contribute to the strong temperature dependence. The strong increase in the current for *V*$_b$>1V even at low temperatures (Fig. 3B inset) suggests over-barrier transport between graphene and WS$_2$.

For devices where BGr was made of monolayer graphene, gating with the Si back gate (through 300nm SiO$_2$) is possible, Fig. 3C. The density of states in monolayer graphene around the Dirac point is very low, which allows manipulation of the work function of graphene and the electric field penetrating to the WS$_2$ barrier with the gate voltage. The zero-bias resistance is not sensitive to the back gate voltage *V*$_g$ applied, whereas the current in the non-linear region demonstrates a 30% modulation when *V*$_g$ is swept between -60V and 60V. The fact that the gate voltage mostly affects the non-linear part of the *I*-*V*$_b$ dependence indicates that the changes in the current are mostly due to the changes in the relative position of the Fermi energy with respect to the top of the valence band in WS$_2$ (as has been previously suggested for tunnelling transistors produced from monocrystalline WS$_2$ (*12*)) and not due to the gating of WS$_2$. The gating of WS$_2$, which would result in modification of the shape of the tunnelling barrier (making it triangular), is not very efficient in LPE samples due to large impurity band (due to edges), which screens the electric field. Note, that from the slope of the conductivity versus *V*$_g$ one can conclude that it is hole transport through the valence band of WS$_2$, which dominates the current, contrary to the conclusion of (*12*), where monocrystalline WS$_2$ has been used. It might be due to the fact that the Fermi level in small flakes of WS$_2$ is pinned close to the valence band by the edge states.

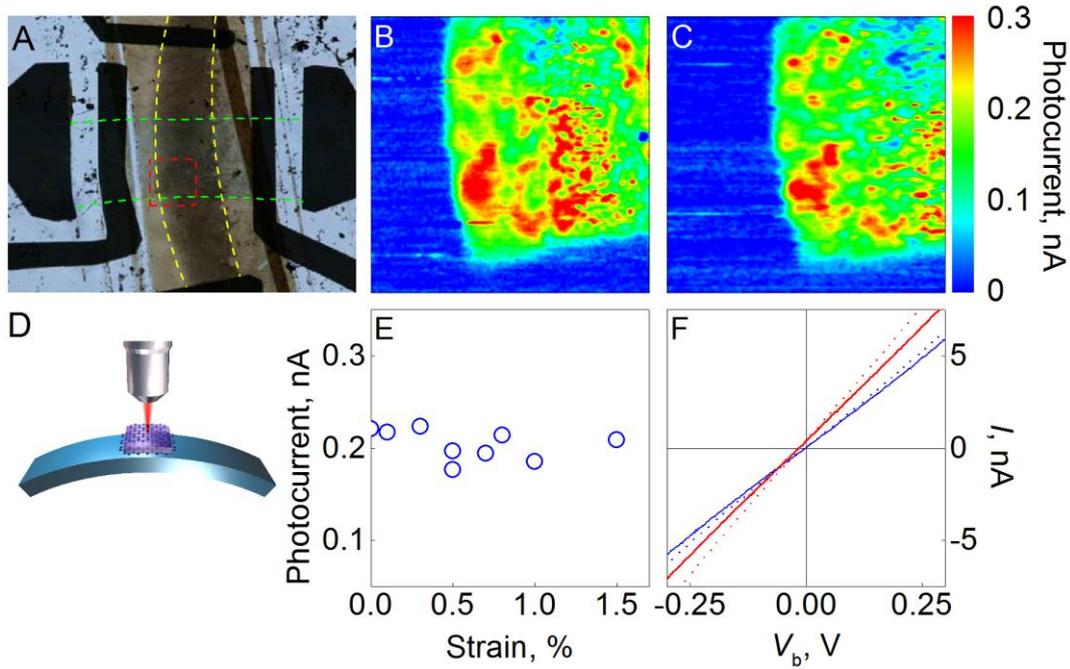

*Figure 5.(A) Optical micrograph of a LPE PET/BGr/WS$_2$/TGr heterostructure. The yellow dotted lines indicate the boundaries of LPE BGr; the green dotted lines - CVD TGr; the red square shows the area investigated by photocurrent mapping (size 70μm×70μm). The brownish stripe which covers the BGr is 60nm LPE WS$_2$. Photocurrent maps (70μm×70μm) taken at an incident power of 190 μW and energy of 1.96 eV at two different curvatures: 0mm$^{-1}$ (B, corresponds to zero strain) and 0.15mm$^{-1}$ (C, corresponds to 1.5% strain). (D) schematic representation of our bending set-up. (E) Average photo-current obtained from the photo-current maps as a function of the applied strain. (F) I-V$_b$ characteristics with (red) and without (blue) illumination for the strained (solid curves) and unstrained (dashed curves) cases. The illumination (power 190 μW) was focused into ~1μm$^2$ spot.*

Similar structures have been used for photovoltaic applications. Again, in the main text we limit ourselves to Si/SiO$_2$/BGr/WS$_2$/TGr type heterostructures (see results for MoS$_2$ in (*25*)). Upon illumination, electron-hole pairs generated mostly in the TMDC layer (due to its high optical absorption(*13*)) can decay into separate electrodes (provided there is an electric field to separate the charges), producing a photovoltage (*13*). Under illumination the *I-V$_b$* characteristics become increasingly linear, Figure 4 (inset), demonstrating that in this regime the current is dominated by the photo-excited carriers. Also, finite photocurrent has been observed even at zero bias voltage (Fig. 4 inset to left panel), demonstrating that such structures can be indeed used as photovoltaic devices.

We recorded zero-bias photocurrent as a function of the position of the laser spot (less than 1μm in diameter) on the device by using a 100x microscope objective. Zero-bias photocurrent maps, taken at different back gate voltages, demonstrate that the photocurrent is produced only in the regions where all three layers (BGr, TMDC and TGr) overlap. We notice that the edges of the sample provide slightly off-set gate dependence of the photocurrent, which we attribute to the effect of environmental doping of the TMDC not covered by TGr. Similar to the case of the transistor, the back gate voltage controls the value and the direction of the electric field across WS$_2$, and thus the magnitude and the polarity of the photocurrent (Fig. 4). For the largest electric field across WS$_2$ (at $V_g$=60V) used in this measurements, we achieved photoresponsivity values of ~ 0.1 mA/W.

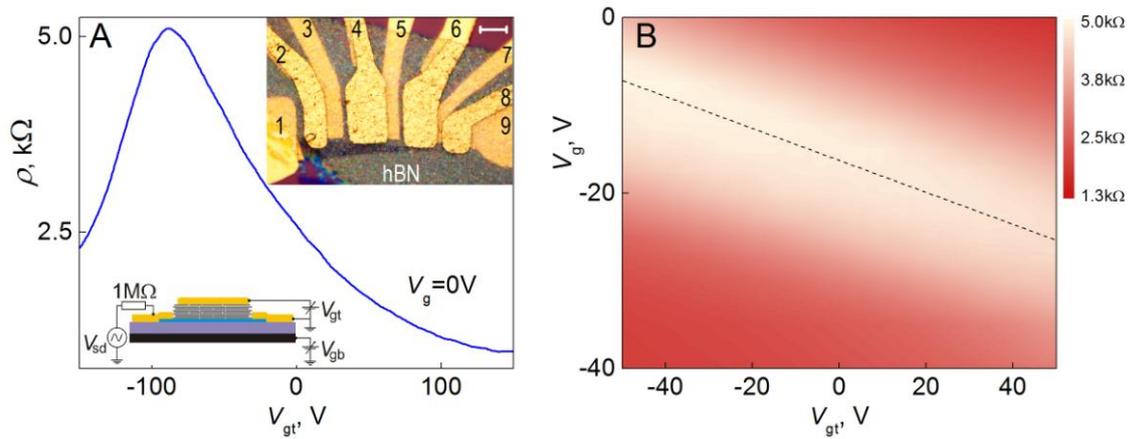

*Figure 6 (A) Resistivity as a function of $V_g$ for one of our Si/SiO$_2$/BGr/hBn/Au device (schematic in the bottom inset). T=300K.Top inset: optical micrograph of one of our device. BGr is contacted by odd-numerated contacts, and gated though LPE hBN (seen as colourful, partly transparent areas) by even-numerated contacts. Scale bar 10μm. (B) Resistivity as a function of $V_g$ and $V_{gt}$. Dashed line indicates the position of the resistivity maximum. T=300K.*

The efficiency could still be increased by using larger flakes, electron and holes scattering and localization on the impurities and edges is reduced, which, in turn, would reduce the contribution of recombination mechanisms(33). We would like to stress that our devices do not require exactly monolayer TMDC to be used, which simplifies the procedure even further. Using even thicker flakes (by reducing the sonication time) means that those would also be larger laterally and allow more efficient charge transfer between the layers, thus allowing for more efficient e-h separation. Also, the use of thick TMDC flakes ensures that the band-structure of TMDC used has a non-direct band-gap(*34, 3*5), thus reducing the probability of recombination. Note, that the photo-absorption for TMDC (per layer) practically does not change with the number of layers(*13*).

Although the photoresponsivity of our devices is significantly smaller than that obtained in current state of the art photovoltaic devices(*36*) or in similar heterostructures based on monocrystalline WS$_2$(*13*), the advantage of our structures is that they can be produced by a variety of low-cost and scalable methods, and are compatible with flexible substrates (note, that the use of CVD graphene as a back electrode for solar cell application is compatible both with the flexible substrates and with this low cost method). To this end we fabricated PET/BGr/WS$_2$/TGr heterostructures on a flexible

PET substrate (thickness 0.2mm) (Fig. 5A). We tested two different methods for sample fabrication: BGr and WS$_2$ layers were produced by either drop-casting or vacuum filtering (with subsequent wet transferring) of the respective LPE dispersion. Both layers were shaped into strips by mechanical removal of the unnecessary material (Fig. 5A). We used CVD graphene as TGr to achieve maximum optical transparency. A 4-point bending rig was utilised to apply uniaxial strain to the heterostructure (Fig. 5D).

As in the previous experiment we scanned a laser across the sample while simultaneously recording the photocurrent, (Fig. 5B,C). The photocurrent is only observed when illuminating the area where all three layers (BGr, WS$_2$ and TGr) overlap. After bending, some local variation in the photocurrent was detected. However, the overall pattern (Fig. 5B, C), the integral value of the photocurrent (Fig. 5E) and the overall resistance of the device (Fig. 5F) remain practically independent of the strain, demonstrating the possibility to use such heterostructures for flexible electronics.

Finally, we demonstrate a different type of heterostructure where LPE hBN is used as a gate dielectric. The dielectric properties of hBN(*19, 29*), added to its excellent chemical and thermal stability, mechanical and thermal properties(*1*), make hBN thin films a promising dielectric alternative in the next generation of nanodevices(*37*). Here we tested Si/SiO$_2$/BGr/hBn/Au devices, where LPE hBN (prepared through filtering of a hBN suspension, with subsequent transfer of the hBN paper from the filter to the device) served as transparent dielectric between the channel (BGr, CVD graphene) and the gate (Au), (Fig. 6A, inset). We also used mechanically exfoliated single and few layer, CVD and LPE graphene as a top electrode.

Resistivity of the BGr channel as a function of top gate voltage $V_{gt}$ is presented in Fig. 6A. The contour plot of the resistivity as a function of $V_g$ and $V_{gt}$ demonstrates the usual resistivity maximum shifting across a diagonal of the plot (Fig. 6B, the dashed line). The slope of the line allows us to establish the ratio of the capacitances to Si and top gate (here we ignore the finite compressibility of 2D electron gas in graphene). Knowing the thickness of hBN from the AFM study (600nm for this particular sample), allows us to estimate the effective dielectric constant of LPE hBN to be ~1.5. The significant deviation from the bulk value (~4, as established in recent tunnelling experiments, (*38*)) is due to loose packing of hBN laminates. This low value of the dielectric constant of the LPE h-BN could be an advantageous property when considering its incorporation in densely packed electronic elements, where loss needs to be kept to a minimum. Indeed air gaps in conventional insulators are deliberately induced to reduce the overall effective dielectric constant (39,40). Knowing the capacitance to the top gate allows us to estimate the mobility of the BGr to be of the order of $3\times10^3$cm$^2$/V·s, which is typical of CVD graphene(*4*). This clearly indicates that deposition of LPE hBN does not deteriorate the properties of graphene. We have also tested the breakdown voltage for our LPE hBN(*25*), which turned out to be 0.25V/nm. This is comparable (or better) than traditionally used ink-jet printed dielectric(*41*) or the dielectric strength of spattered films(*42*). This demonstrates that LPE hBN can be used as a dielectric for transparent, flexible transistor applications.

CONCLUSIONS

Demonstrated examples show that inks based on 2D atomic crystals are suitable for printable, flexible and transparent electronics. Moreover, the combinations of different inks allow for the creation of complex heterostructures, which might be suitable for multifunctional applications. Although many of the heterostructures created still underperform in comparison with the benchmark structures, their versatility, low cost, the simplicity of the technology and unique properties (e.g. flexibility and transparency) might prove beneficial for some types of devices. We foresee that multifunctional applications might gain the most, as a large number of very different 2D crystals could be combined in one stack. At the same time, some applications, such as the use of hBN as high performance dielectric material, are already in the mature state. Furthermore, the possibility of fine-tuning the properties of the inks, by varying the size and thickness of the flakes as well as the type of solvent – will increase the range of functionalities of the resulting heterostructures and devices even further.

MATERIALS AND METHODS

The dispersions were produced via sonication of the original crystals in different solvents (NMP, water and water/ethanol mixture). The films were produced by drop-casting, inkjet printing and filtering and were characterised by Raman spectroscopy, AFM, TEM, and XPS. Photocurrent measurements were performed by irradiating the active area of the devices with a laser of energy 1.96 eV. A 100x objective with numerical aperture 0.60 was used to focus the spot to less than 1 µm diameter. The power was accurately measured using a Thorlabs PM100 power meter with sensitivity 10 nW. A voltage drop over a known resistor placed in series with the photodiode was used to calculate the photocurrent. We then utilized a piezo stage with accuracy 100 nm to move the sample under the laser beam. The stage position was linked to the photocurrent measurements in order to obtain spatial information of the photocurrent. The external quantum efficiency (EQE) is calculated as the number of charge carriers collected by the contacts to the number of incident photons, $EQE = \frac{hf}{e} \frac{i_{pc}}{P}$, where h is Plank's constant, f is the frequency of the incident photon, e is the electron charge, $i_{pc}$ is the photocurrent at zero bias and $P$ is the incident laser power (See (*25*) for detailed analysis of the photovoltaic performance).

Acknowledgments

This work was supported by The Royal Society, U.S. Army, European Science Foundation (ESF) under the EUROCORES Programme EuroGRAPHENE (GOSPEL), European Research Council and EC under the Graphene Flagship (contract no. CNECT-ICT-604391). Y.-J. Kim's work was supported by the Global Research Laboratory (GRL) Program (2011-0021972) of the Ministry of Education, Science and Technology, Korea. F. Withers acknowledges support from the Royal Academy of Engineering; A. Felten is a FRS-FNRS Research Fellow.

Supporting Information Available: detailed information on material preparation, characterization and fabrication of devices, including additional devices are provided. This material is available free of charge via the Internet at http://pubs.acs.org.